\documentclass[a4paper,fleqn, authoryear]{cas-dc}

\usepackage[]{natbib}
\usepackage{amsmath}
\usepackage{bm}
\usepackage[utf8]{inputenc} 
\usepackage[T1]{fontenc}    
\usepackage{graphicx}
\usepackage{hyperref}
\usepackage{url}            
\usepackage{booktabs}       
\usepackage{amsfonts}       
\usepackage{nicefrac}       
\usepackage{microtype}      
\usepackage{lipsum} 
\usepackage{bm}
\usepackage{subcaption}
\usepackage{caption}
\usepackage{xcolor}

\def\tsc#1{\csdef{#1}{\textsc{\lowercase{#1}}\xspace}}
\tsc{WGM}
\tsc{QE}
\tsc{EP}
\tsc{PMS}
\tsc{BEC}
\tsc{DE}

\begin{document}
\let\WriteBookmarks\relax
\def\floatpagepagefraction{1}
\def\textpagefraction{.001}
\shorttitle{}
\shortauthors{}

\title [mode = title]{Beyond Stationary Simulation; Modern Approaches to Stochastic Modelling}
\tnotemark[1]
\tnotetext[1]{Research funded by GoldSpot Discoveries Corp.}





                        
\author[1]{Pejman Shamsipour}[]
\ead{pejman@goldspot.ca}
\fnmark[1]
\author[3]{Tedd Kourkounakis}[orcid=0000-0003-2292-5429]
\ead{tedd.kourkounakis@goldspot.ca}
\fnmark[1]
\cormark[1]
\author[2]{Amin Aghaee}[]
\ead{amin@goldspot.ca}
\fnmark[1]
\author[3]{Rouzbeh Meshkinnejad}[]
\ead{rouzbeh.meshkinnejad@goldspot.ca}
\fnmark[1]
\author[3]{Manit Zaveri}[]
\ead{manit.zaveri@goldspot.ca}
\fnmark[1]
\author[2]{Shawn Hood}[]
\ead{shawn@goldspot.ca}
\fnmark[1]


\address[1]{GoldSpot Discoveries Inc., 980 Rue Cherrier, Montréal, QC H2L 1H7}
\address[2]{GoldSpot Discoveries Inc., 303- 700 West Pender, Vancouver, BC, V6A 1V7}
\address[3]{GoldSpot Discoveries Inc., 69 Yonge St \#1010, Toronto, ON M5E 1K3}


\cortext[cor1]{Corresponding author}
\fntext[fn1]{All authors share equal contribution towards this work.}


\begin{abstract}
Stochastic and conditional simulation methods have been effective towards producing realistic realizations and simulations of spatial numerical models that share equal probability of occurrence. Application of these methods are valuable throughout the domain of earth science for their ability to simulate sampled study data. Such stochastic methods have also been adopted into other fields outside of geostatistics domains, especially within the computing and data science community. Classical techniques for stochastic simulation have primarily consisted of stationary methods due to their brisk simulation speed and mathematical simplicity. However, advances in modern computing now allow for the implementation of more advanced non-stationary simulation methods, consisting of multiple varying structures, and allowing for much more accurate and realistic simulations. As some of these calculations may still be slow, the application of machine learning techniques, namely the Generative Adversarial Network (GAN) can be used to surpass previous simulation generation speeds and allows for greater parameterization. This work presents three stochastic simulation methods: stochastic simulation using non-stationary covariance, multipoint simulation, and conditional GANs. An SPDE method was used as a benchmark comparison. Experiments using synthetic data are used to showcase the effectiveness of each of these methods at maintaining non-stationary structures and conditioned data. A case study implementing non-stationary covariances is also presented on real geochemical samples coming from La Ceinture de roches vertes de la Haute-Eastmain, located in the Superior Province. 



\end{abstract}

\begin{graphicalabstract}
\includegraphics{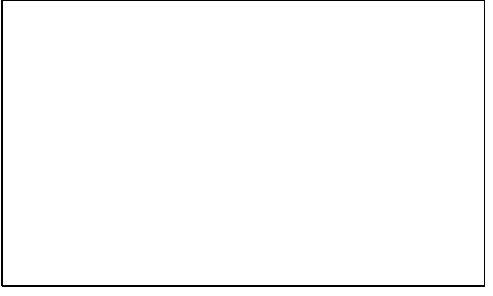}
\end{graphicalabstract}

\begin{highlights}
\item Stochastic simulation using non-stationary covariance is presented
\item Multipoint simulation is experienced
\item Simulation using cGAN is carefully tested
\end{highlights}


\maketitle

\section{Introduction}
\label{section:Introduction}
Stochastic simulation models have been effective towards producing realistic realizations and simulations of numerical models that share equal probability of occurrence. Application of these simulation models have been used throughout many domains of earth science \citep{chiles2009geostatistics} and have recently expanded into other fields, including computer vision, machine learning \citep{rasmussen2003gaussian}, among others. 

Classical or traditional geostatistical and Gaussian process models are designed with the assumption of stationary covariance. This means that the covariance function used is assumed to be invariant over the entire study area, and implies spatial homogeneity of the data within this area. This suggests that the main geologic features share some similarities in size or spatial continuity (covariance function correlation range), orientation (covariance function main anisotropy directions), and variability of the studied property (variance) throughout the entirety of the study area \citep{emery2018continuous}. This stationarity assumption, however, is not omnipotent. In cases of non-uniform geological structures, for example, where some subdomains present different preferential geologic directions, or when the sizes of the underlying geologic structures vary significantly in subareas or along a given direction, stationary algorithms struggle to effectively model these areas \citep{chiles2009geostatistics}. Geologic contexts showing faults and dikes along different directions for separate sub-areas are unlikely to be well described by a single covariance model. Similarly, large folds are expected to show different preferential directions along their limbs. A known regional geologic discontinuity (e.g., fault or erosional surface) can separate the study area into sub-domains with different statistical properties.

Bayesian inversion algorithms have successfully been applied to geophysical inversions. In most of these techniques, model covariance parameters are assumed to be stationary. In order to take into account the petrophysical variability result from inversion, \citet{shamsipour20133d} suggest using non stationary covariances.

To better represent these multi-structured areas, this work describes the implementation of non-stationary covariance functions. Non-stationary models allow for the point-wise consideration of locally varying orientations and ranges caused by the aforementioned geological phenomena. They may inherently act as a stationary model given constant structure definitions throughout a study area, but may also adapt to modest or sudden changes across a given region. The study of non-stationary modelling has grown recently within multiple domains, including that of geosciences, with new methods continuously being explored (\cite{emery2018continuous}, \cite{fouedjio2017second}, \cite{risser2016nonstationary}, \cite{fouedjio2016generalized}). 

Traditional techniques for simulation of complex surfaces implement covariance based two-point statistics which have limited scope to inherit the non-stationary complexities in the mathematically structured realizations. Recent decades have seen continued improvement of the algorithms using Multi Point statistics (MPS), and proved their strong performance in a variety of testing and implementation situations. These MPS still confine to the limitation of stationary data, but with some secondary data, they can achieve non-stationary realizations as well as implement hard and soft conditioning based on the data.      

MPS simulation was developed with an aim to handle complex structural patterns by calculating higher order statistics of the complex features and patterns from training images (TI). The idea of using TI and the probability distribution function is based on the single normal equation proposed by  \citet{guardiano1993multivariate}. That single normal equation is defining a conditional probability using Bayes theorem. With the pioneering work on a single normal equation, important advances on practical applications have been developed. Extending the application of the single normal equation, the SNESIM algorithm was developed by \citep{strebelle2002conditional} which popularized the use of MPS for generating simulated images. Even though the original SNESIM stands in need of input training image to be fairly stationary, new extended versions of the idea can account non stationary constraint as well (\cite{wu2007non}, \cite{zhang2006filter}). Other popular MPS simulations algorithms that handle non-stationarity include Direct Sampling (DS) \citep{mariethoz2010direct} as well as Cross-correlation simulation (CCSIM) \citep{tahmasebi2012multiple} which focus on optimizing the computation time. Limiting the option of scalability is witnessed in 3D process based models. A more comparative study along with advantages, disadvantages and challenges of several MP models is provided by \citet{tahmasebi2018multiple}.

One shortcoming of non-stationary models are their significantly longer covariance function calculation time. These methods have a time complexity of $O(n^{3})$. Extremely high resolution outputs take exceedingly long amounts of time, especially three-dimensional simulations. As an alternative to non-stationary covariances, this work also proposes the implementation of a modern deep learning model, based on the GAN \citep{goodfellow2014generative}. The GAN, alongside its multiple variations (CycleGAN \citep{zhu2017unpaired}, StackGAN \citep{zhang2017stackgan}, among others) have been applied to a wide breadth of domains for the purpose of data synthesis and simulation. They can be used to bypass prolonged calculations and let the model innately learn a given structure provided synthetic representations of an area are available. After training the network, conditioned or unconditioned simulations can be generated significantly quicker than the classical methods. We use this model to generate simulated images that mimics structures shown in input training data, as done by \citet{dupont2018generating, song2021gansim}. 

\textcolor{black}{In this paper, we compare three stochastic simulation methods, namely a stochastic simulation using non-stationary covariance, multipoint simulation, and conditional generative adversarial network (cGAN) against the benchmark Stochastic Partial Differential Equation (SPDE) method.}

The key elements of our work can be summarized as follows:
(i) The implementation of multi-dimensional non-stationary covariances, alongside a comparison to the stationary approach;
(ii) A showcase of the method and implementation of multipoint simulation for non-stationary structures;
(iii) Utilizing generated samples from the non-stationary covariance and multipoint simulations, we implement a Generative Adversarial Network to simulate non-stationary structured areas.
Note that each of these methods may also incorporate conditioned simulations on provided input data.
(iv) A discussion and comparison of the aforementioned implementations to existing non-stationary models.

The rest of this paper is organized as follows; Section \ref{section:Methodology} contains a breakdown of the procedures and algorithms of non-stationary, multipoint, and GAN simulation methods. Section \ref{section:Experiments} will define the conducted experiments in further detail. A case study for implementing non-stationary covariances can be found in Section \ref{section:Case Study}, followed by a discussion of the results and conclusion in the final section.

\section{Methodology}
\label{section:Methodology}
\subsection{Non-stationary covariance}
\textcolor{black}{\cite{higdon1999non}} first introduced \textcolor{black}{on-stationary} covariance functions, derived from convolving kernel functions:

\begin{equation} 
C^{ns}(\bm{d}_i, \bm{d}_j)=\int_{\mathbb{R}^d} K_{d_i}(\bm{u})K_{d_j}(\bm{u})\mathrm{d}\bm{u}
\label{eq:higdon}
\end{equation}

where $d_i$, $d_j$ and $u$ are locations in $\mathbb{R}^d$, and $K_{d_i}$ is a kernel function centered at location $d_i$.
\citep{paciorek2006} proved that this covariance function is positive definite in all Euclidean spaces.

When the kernels are Gaussian, $K_{d_i}(.)$ in Eq. \ref{eq:higdon} represents a multivariate Gaussian density distribution, centered on $d_i$ with covariance matrix $\Sigma_i$. Using convolution, the non-stationary covariance function between any pair of points $d_i$ and $d_j$ therefore becomes:
\begin{equation} 
C^{ns}(\bm{d}_i, \bm{d}_j)=\sigma^2 |\Sigma_i|^{\frac{1}{4}} |\Sigma_j|^{\frac{1}{4}} |\frac{\Sigma_i+\Sigma_j}{2}|^{-\frac{1}{2}}exp(-D_{ij})
\label{eq:paciorek}
\end{equation}
where $\sigma^2$ is variance and
\begin{equation}
 D_{ij}=|(d_i-d_j)^T (\frac{\Sigma_i+\Sigma_j}{2})^{-1}(d_i-d_j)|.
\label{eq:D}
\end{equation}
as shown by  \citep{paciorek2003nonstationary}.

For the stationary case, $\Sigma_i=\Sigma_j=\Sigma$ and, as a result, $D_{ij}$ become the Mahalanobis square distance between vectors $d_i$ and $d_j$. However, in the non-stationary case, $D_{ij}$ is seen as a `mean' Mahalanobis square distance between these two points.  

This covariance is infinitely differentiable at the origin, which may not be desired in many applications. A more general non-stationary covariance model has been proposed by \citet{paciorek2006} as:
\begin{equation} 
R^{ns}(\bm{d}_i, \bm{d}_j)= |\Sigma_i|^{\frac{1}{4}} |\Sigma_j|^{\frac{1}{4}} |\frac{\Sigma_i+\Sigma_j}{2}|^{-\frac{1}{2}}R^s(\sqrt {D_{ij}}).
\label{eq:paciorekg}
\end{equation}
where $R^{s}$ is an isotropic correlation function which is positive definite on $\mathbb R^p$ for every $p$ in \{1, 2, $\hdots$.\}.
The exponential covariance model is a member of this class \citep{chiles2012generalized} and can be generated using (Eq. \ref{eq:paciorekg}) where $D_{ij}$ is the 'average' Mahalanobis distance (Eq. \ref{eq:D}) for every pair of points.


These final non-stationary covariances can be applied for the inversion of potential field data, as described by \citet{shamsipour20133d}.

Conditional simulation provides an ensemble of realizations of the random function. These realizations are carefully conditioned upon the data. Note that non-linear statistics can be computed from these ensemble simulations. We can condition our simulations (including non-stationary) using the following equation \citep{chiles2009geostatistics}:

\begin{equation}
Z_{cs}(x) = Z^*(x) + (Z_s(x)-Z^*_s(x))
\label{eq:conditional}
\end{equation}
where:\\
$Z_{cs}$: Conditional Simulation\\
$Z^*(x)$: Kriged data\\
$Z_s(x)$: Simulated data\\
$Z^*_s(x)$: Kriged simulated data\\

The term, $(Z_s(x)-Z^*_s(x))$, of Equation \ref{eq:conditional} is called simulated kriging error in some literature \citep{wackernagel2003multivariate}.

\subsection{Multipoint simulation}

In this paper, we are specifically targeting the methods that can simulate non-stationarity patterns in 2D and 3D facies. After careful review of available MP methods, we implement a specific MPS method, `SNESIM', because of its stability and popularity. As MP methods are very popular amongst the geological domain, we summarize the workflow of the algorithm from the original manuscript provided by the author in \citet{strebelle2002conditional}. 

This algorithm falls in the category of pixel-based algorithms, which can perfectly reproduce the conditioning point data. The generated model also stores all of the frequency/probability of the complex patterns from the TI and saves them in a search tree, significantly reducing computation time \citep{boucher2009considering}.

\begin{enumerate}
    \item Assign the original sample data to the closest grid nodes. Define a random path based on the distribution of the data nodes.
    \item If the conditioning data falls within the maximum search template, the probability values are retained from the search tree for those unsampled data nodes, $u$. 
    A simulated s-value for that specific node, $u$, is drawn using the search tree and added to the s-data, which is further on used for conditioning and simulation at all subsequent nodes. This step is repeated for all the upcoming nodes from the random path.
    \item Following these steps, one stochastic image will be generated which looks identical to the TI along with similar categorical proportions. Iterating this process with random path generates a new realization. 
\end{enumerate}

The work discussed in this paper extensively depends on the algorithm's proven capability of simulating complex patterns in multi-dimensional geometry. As a result, the realizations developed using SNESIM can be used as a inputs, or training data, for the GAN model. 

\subsection{Generative adversarial network}
To add another benchmark to the previous models, and to improve upon the slower speeds and accuracy of the non-stationary simulation, we designed a GAN \citep{goodfellow2014generative} to conditionally simulate our non-stationary input data. Generative modelling focuses on learning the regularities and trends within an input dataset in order to generate further samples that are indistinguishable from the original given data. GANs have been used in this manner to generate new facial images, fill in missing data, as well as other geological implementations. The use of GANs in this work is to simulate plausible geophysical or geochemical trends corresponding to those of a given dataset, while also being conditioned to known sample points.

A GAN consists of two models: the generator, $G$, and discriminator, $D$. The generator is the network that attempts to generate the desired output data that mimics the provided input data given some noise, $z$. The discriminator is a classifier that learns the representations of the real input data, $x$, and distinguishes between the real and generated samples. The loss functions for these models can be described as:
\begin{equation}
L_{G} = min[log(D(x)) + log(1 - D(G(z)))]
\label{eq:GAN_loss_G}
\end{equation}
\begin{equation}
L_{D} = max[log(D(x)) + log(1 - D(G(z)))]
\label{eq:GAN_loss_D}
\end{equation}

\begin{figure}
    \centering
    \includegraphics[width=.5\textwidth]{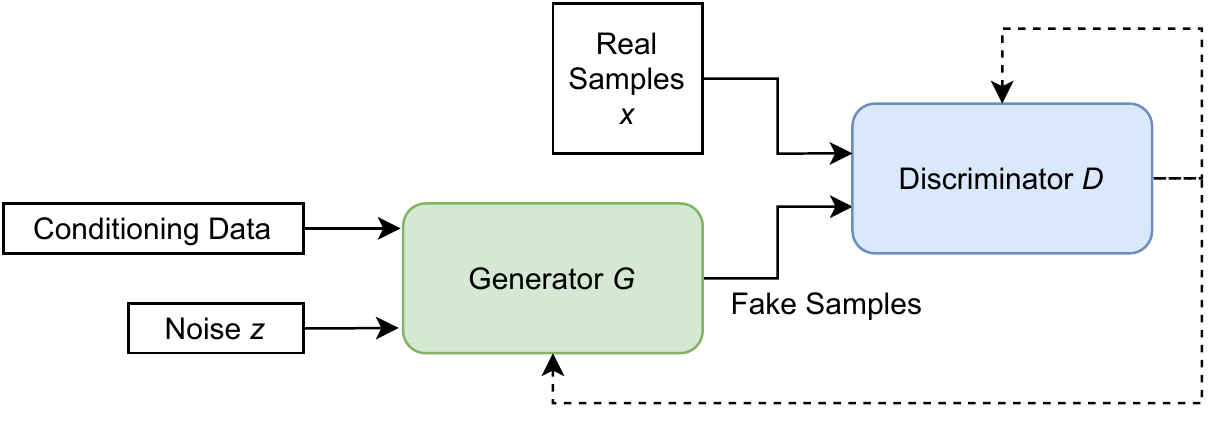}
    \caption{General workflow of the Generative Adversarial Network.}
    \label{fig:GAN}
\end{figure}

Note the discriminator loss in Eq. \ref{eq:GAN_loss_D} should be maximized, whereas the generator loss in Eq. \ref{eq:GAN_loss_G} is to be minimized. These interconnected models are trained simultaneously, with the loss of the discriminator affecting how the generator develops new data. This workflow can be seen in Figure \ref{fig:GAN}.

GAN simulations can simply be conditioned by adding the context loss. This loss penalizes differences between generated samples and known pixels, $y$. Suppose $M$ is the masking matrix, the context loss can defined as follows \citep{yeh2016semantic, song2021gansim}:

\begin{equation}
L_{C}(z|y, M) = ||M \odot (G(z)-y))||_1
\label{eq:GAN_loss_Context}
\end{equation}
where $\odot$ represents component-wise multiplication for matrices.

\subsection{Stochastic Partial Differential Equation}
Gaussian Fields (GFs) are fundamental in spatial statistics particularly in Geostatistics. Even though GFs are well suited to both an analytical and a practical perspective, the computation time is often a drawback. This is caused by the time complexity of $O(n^{3})$, largely due to the calculations of covariance matrices. Lindgren et al., 2011 introduced the Stochastic Partial Differential Equation (SPDE) for the first time for generating Gaussian fields. The idea is summarized in these 3 steps (\cite{lindgren2011explicit}):

Firstly, generating a discretized GF with a corresponding matrix $\Sigma$ by utilizing a GF on a set of locations. Next, use the generated GF to discover corresponding representative generate a discretely indexed Gaussian Markov Random Field (GMRF) with precision matrix Q. These two processes allow us to create a GMRF with a sparse precision matrix, while also having an explicit link to the original GF. Finally, we can conduct all of the calculations using the GMRF. This results in faster calculations whilst still using the modelling of the GF. It should be noted that the SPDE approach can simply be adapted for non-stationary and anisotropic random fields as well.

Though an effective method for non-stationary stochastic modelling method, it does come with some drawbacks. Firstly, utilizing SPDE restricts the each solution for a sample on the random fields to require a linear system. Also, the correlation between random fields is innately defined through the PDE itself. SPDE and similar methods also come with a notable preprocessing time costs. On the negative side, the approach comes with an implementation and preprocessing time costs for  setting up the models, and though minimized with modern computing technology, such time sinks are unavoidable with large calculations.

\section{Experiments}
\label{section:Experiments}

To demonstrate the uses of non-stationary covariances, we calculate these covariances for both one and two dimensional data for resource estimation and simulation. This work also includes two other conditioned simulation methods used as benchmarks: multipoint simulation and GANs. Note that although this work only applies up to three dimensional covariances, these applications can be easily be further scaled to greater dimensions of data, exponentially scaling the result generation time.

\subsection{Non-Stationary covariance}
\subsubsection{1D simulation}
\begin{figure}
    \centering
    \includegraphics[width=.5\textwidth]{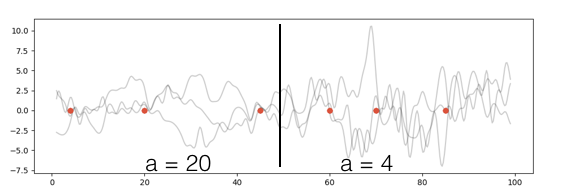}
    \caption{Conditional realizations for 1D non-stationary simulation.}
    \label{fig:1d}
\end{figure}

Figure \ref{fig:1d} shows an example of three realisations of non-stationary Gaussian process regression using a spherical model with a length distance change occurring at $x = 50 m$. These realisations can also be obtained by conditional simulation techniques using kriging \citep{chiles2012generalized}. All of these these realizations have also been conditioned with a series of points with a value of zero, as shown by the red scatter points. Whereas a stationary model would only allow for a single all encompassing length distance, this non-stationary covariance function allows for a clear difference in range across each side of the data.

\subsubsection{2D simulation}

\begin{figure}
     \centering
     \begin{subfigure}[b]{0.3\textwidth}
         \centering
         \includegraphics[width=\textwidth]{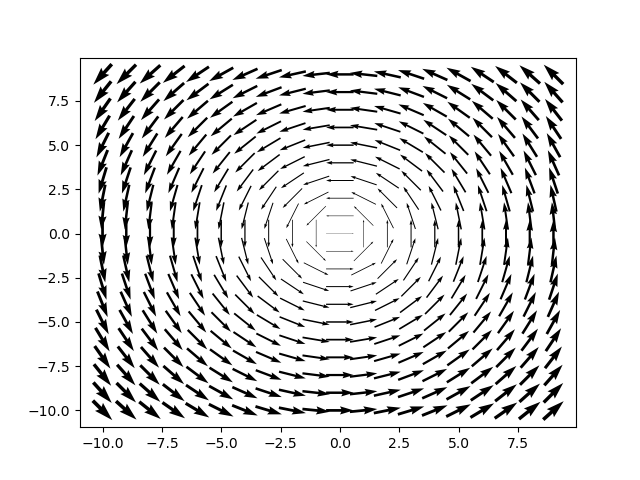}
         \caption{Spiral structure}
         \label{fig:spiral_structure}
     \end{subfigure}
     \hfill
     \begin{subfigure}[b]{0.3\textwidth}
         \centering
         \includegraphics[width=\textwidth]{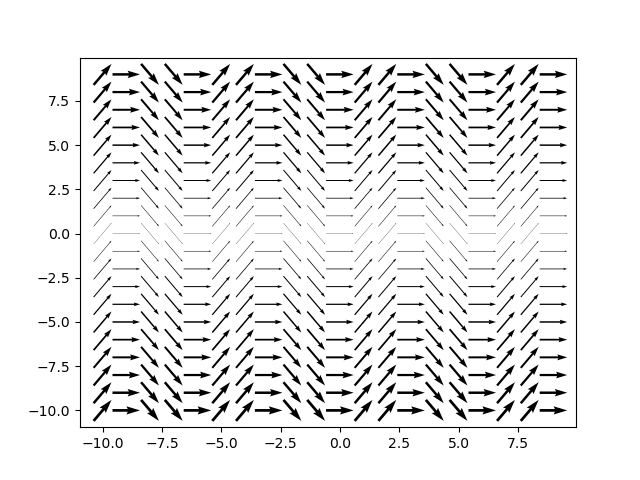}
         \caption{Sinusoidal structure}
         \label{fig:sine_structure}
     \end{subfigure}
        \caption{Orientation and ranges of 2D synthetic simulation structure data.}
        \label{fig:2d_synthetic_structure}
\end{figure}

To demonstrate the effectiveness of the non-stationary simulations, two synthetic structures were generated; the first structure demonstrating a spiral orientation around the midpoint with increasing ranges away from this midpoint, and a second structure represented by a sinusoidal wave along one axis, and an increasing range away from the equator. These orientation and range structures are shown in Figure \ref{fig:2d_synthetic_structure}, where the direction of the arrows represent orientation, and the widths represent range.

\begin{figure}
     \centering
     \begin{subfigure}[b]{0.5\textwidth}
        \centering
        \includegraphics[width=\textwidth]{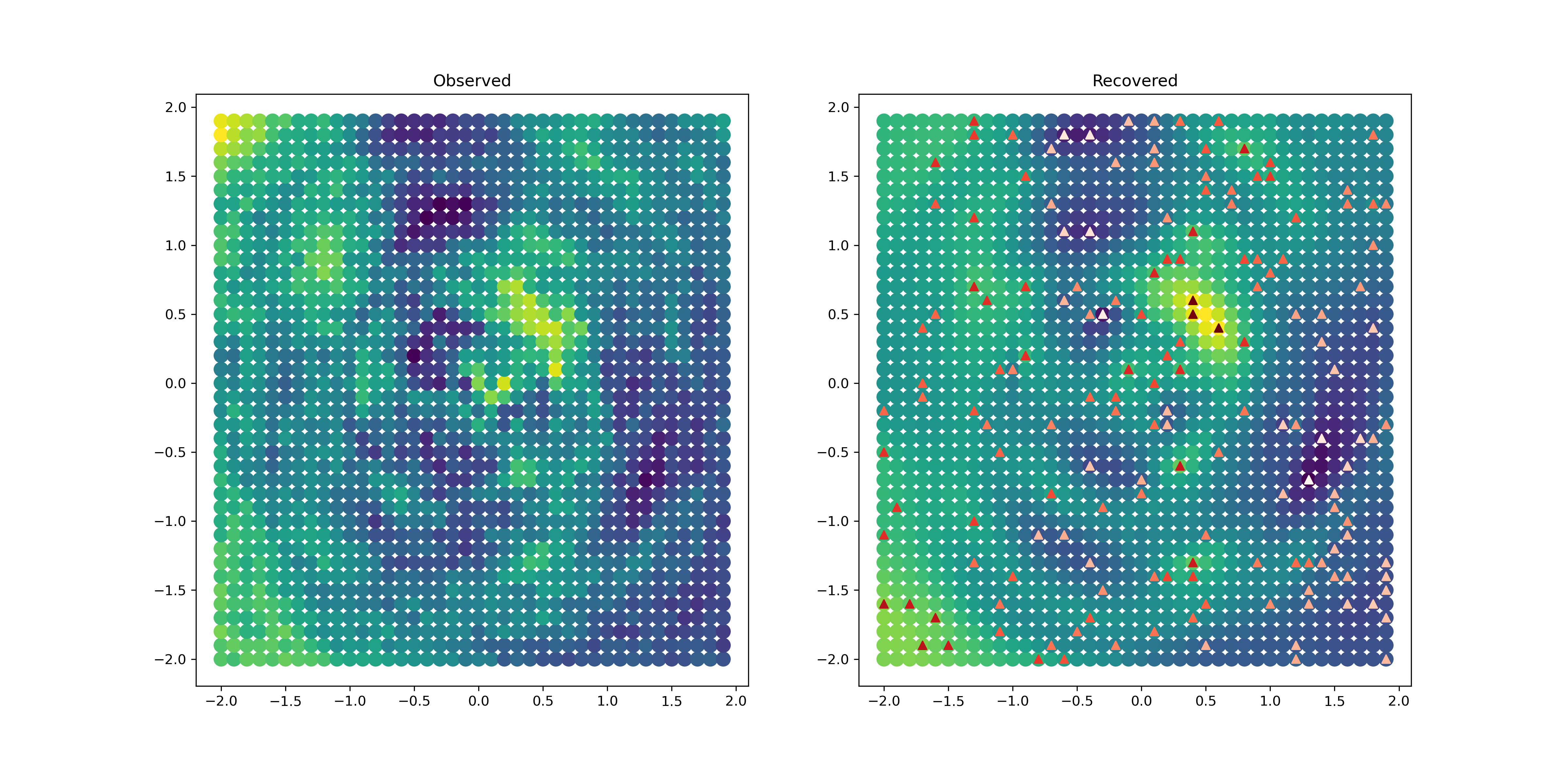}
        \caption{\textcolor{black}{Left: Example non-stationary simulation. Right: Sample conditioning points (range of white to red points corresponding with range of blue to yellow simulation data, respectfully) applied to an example non-stationary simulation.}}
        \label{fig:nonstat}
     \end{subfigure}
     \hfill
     \begin{subfigure}[b]{0.5\textwidth}
         \centering
         \includegraphics[width=\textwidth]{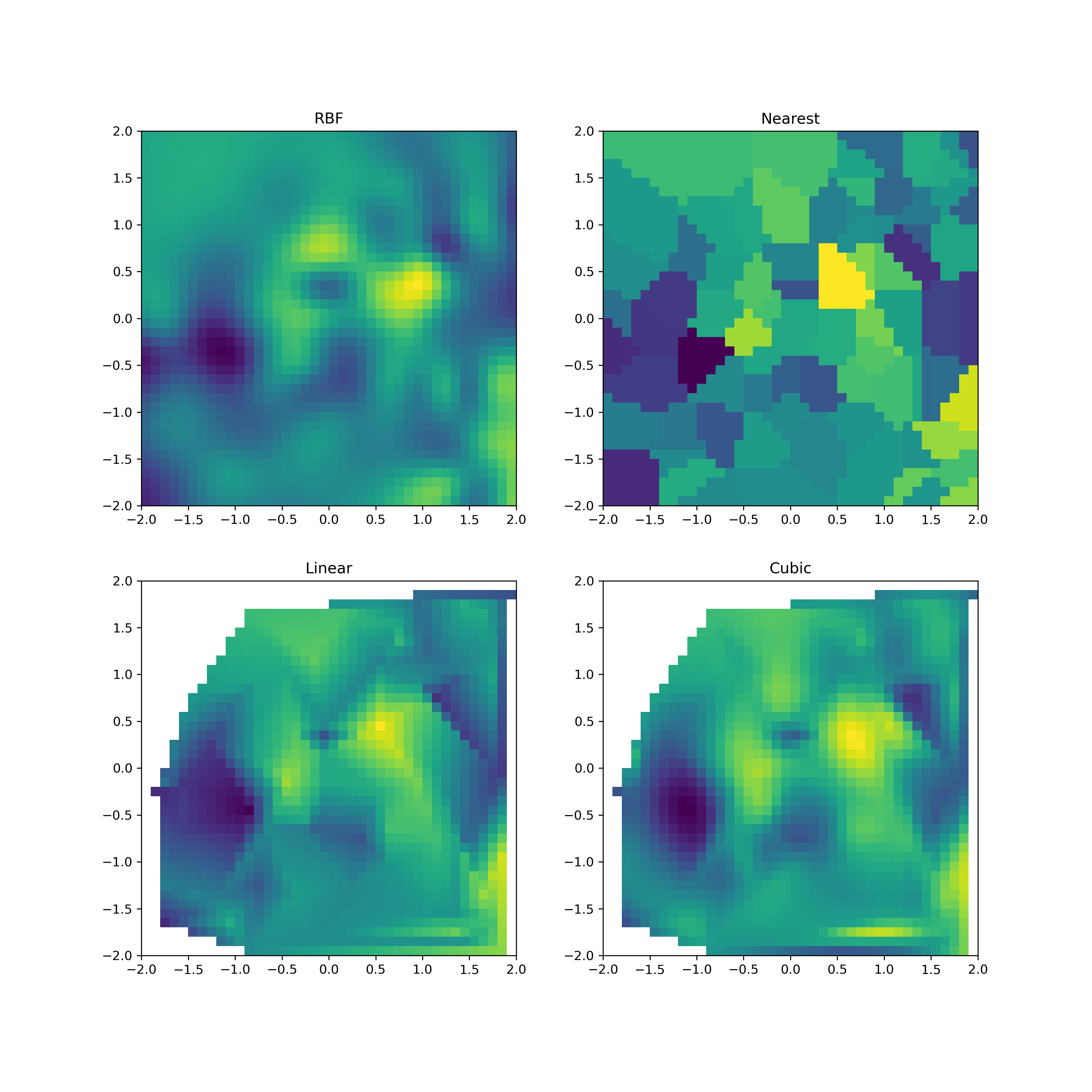}
         \caption{\textcolor{black}{Stationary comparisons on same the conditioning points as Figure \ref{fig:nonstat} (from left to right, top to bottom): Radial basis function; Nearest neighbour; linear interpolation; cubic interpolation.}}
         \label{fig:stat}
     \end{subfigure}
        \caption{Comparison of the proposed non-stationary covariance comparing other stationary methods.}
        \label{fig:2D}
\end{figure}

\textcolor{black}{To demonstrate conditioning on the 2D non-stationary covariances, assume a set of data over a 2D domain with variable orientation (rotation of axes of anisotropy) depicted in Figure \ref{fig:nonstat}. Figure \ref{fig:2D} shows comparison between non-stationary models (\ref{fig:nonstat}) and stationary estimation methods (\ref{fig:stat}) on the same set of conditioning points.} We used Scipy \citep{2020SciPy-NMeth} interpolation techniques, a collection of mathematical algorithms and convenience functions, for generating a variety of estimation examples. These methods include the radial basis function (RBF), nearest neighbour, as well as linear and cubic interpolation methods. In these cases, the mean orientation is reproduced, but in the non-stationary case the local orientation information is better portrayed at its corresponding location. This trend is identical for the range as well. Notably, the spiral structure of the original simulation is maintained in the non-stationary conditioning, but is lost in the other approaches.

\subsection{Multipoint simulation}

For the multipoint simulation experiments, training images from \citet{strebelle2002conditional} are used, consisting of 250 by 250 binary pixel images. For the three dimensional case, a sample training image representing hydrofacies in an alluvial aquifer from Maules Creek, Australia is used \citep{mariethoz2014multipoint}. A 40 pixel cubic sample of this TI was used for training the 3D model. One thousand simulations were generated for both the two and three dimensional experiments. Simulations from this dataset are used to train the 3D GAN, as described in the following subsection.

\begin{figure*}
    \centering
    \includegraphics[width=0.7\textwidth]{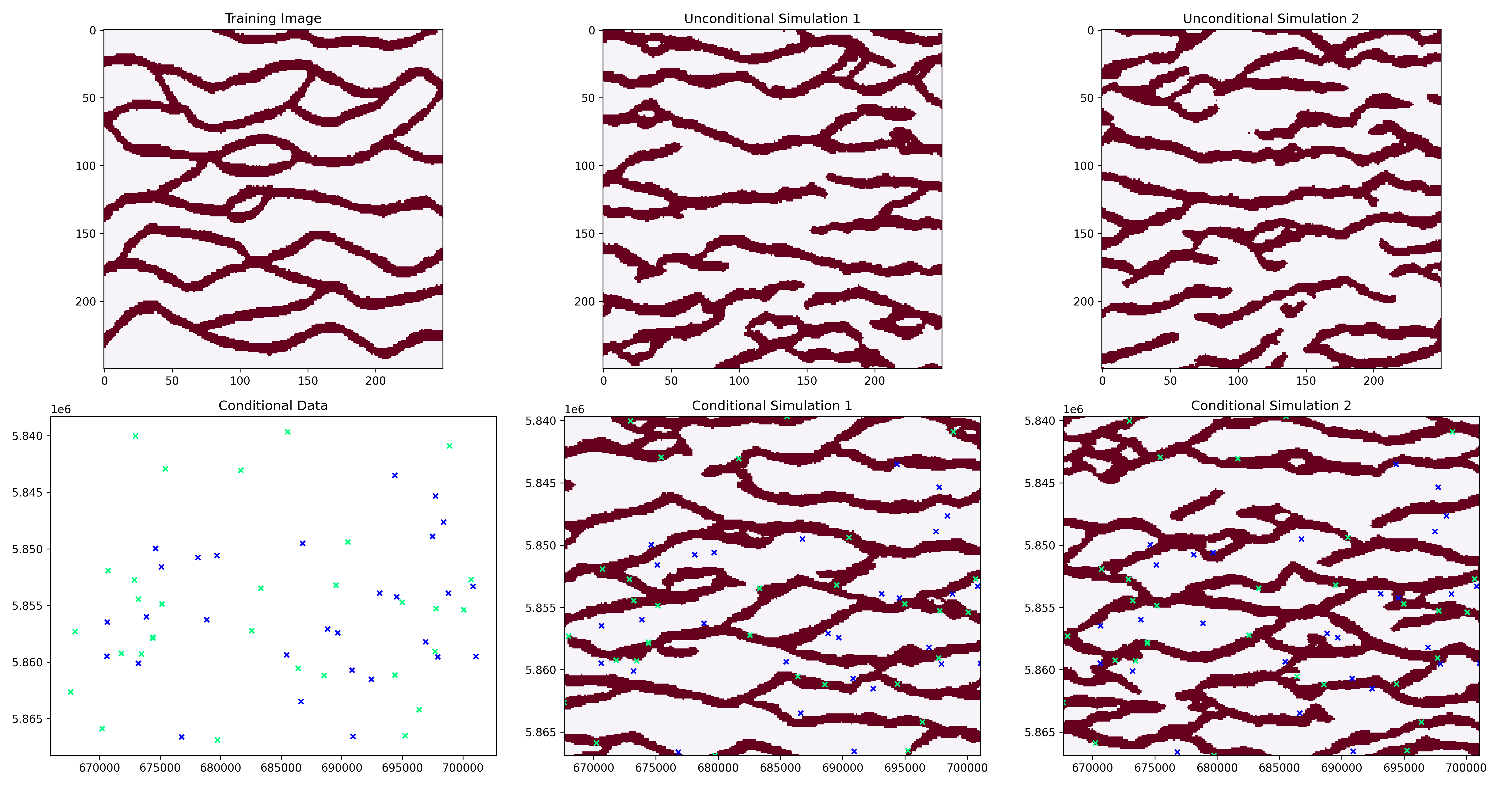}
    \caption{\textcolor{black}{Top: Binary (red and rose channels) Multipoint Strebelle reference input and unconditional simulation results. Bottom: Conditioning points (green and blue points corresponding to red and rose channels, respectively) and conditioned multipoint simulation}}
    \label{fig:2d_multipoint_results}
\end{figure*}

\textcolor{black}{A sample training image, set of conditioning points, and corresponding generated simulations can be found in Figure \ref{fig:2d_multipoint_results}. As exemplified in the top row, the generated simulations maintain the same overall structure as their comparison training image samples. When applying a set of points for conditioning, nearly all provided points are conditioned correctly, while still preserving the original structure of the input samples.}

\begin{figure}
    \centering
    \includegraphics[width=.5\textwidth]{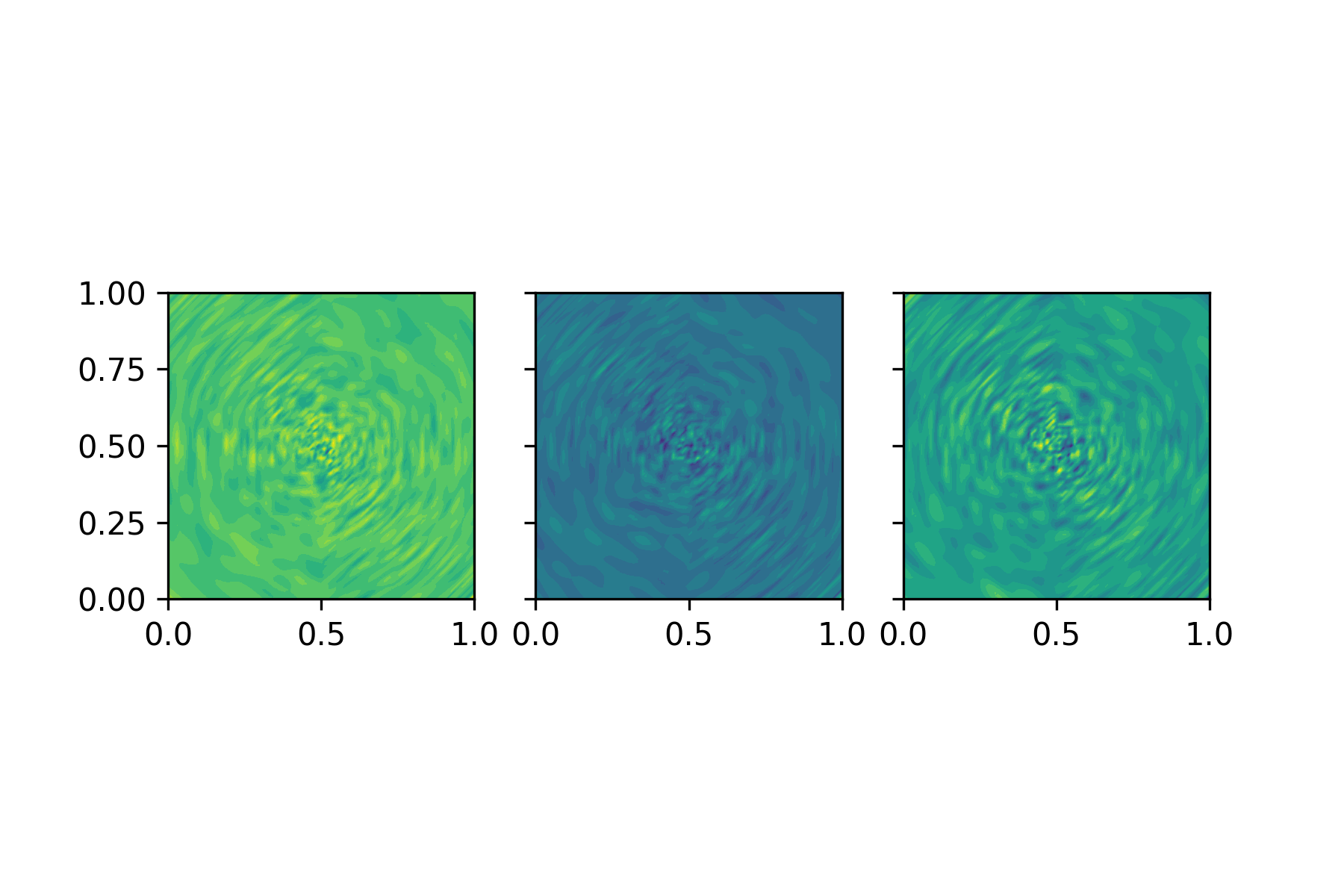}
    \caption{Running three simulations using stochastic partial differential equation}
    \label{fig:2d_sphiral_spde}
\end{figure}

\subsection{Generative adversarial network}
Two pairs of GAN generators and discriminators were trained using our synthetic spiral and sine wave simulations, generated using our non-stationary covariance method. Each model consisted of 1000 simulation samples for training. The models were trained with a learning rate of 0.01 with 700 iterations for the 2D GAN, and 60 iterations for the 3D model.

\begin{figure*}
    \centering
    \includegraphics[width=0.8\textwidth]{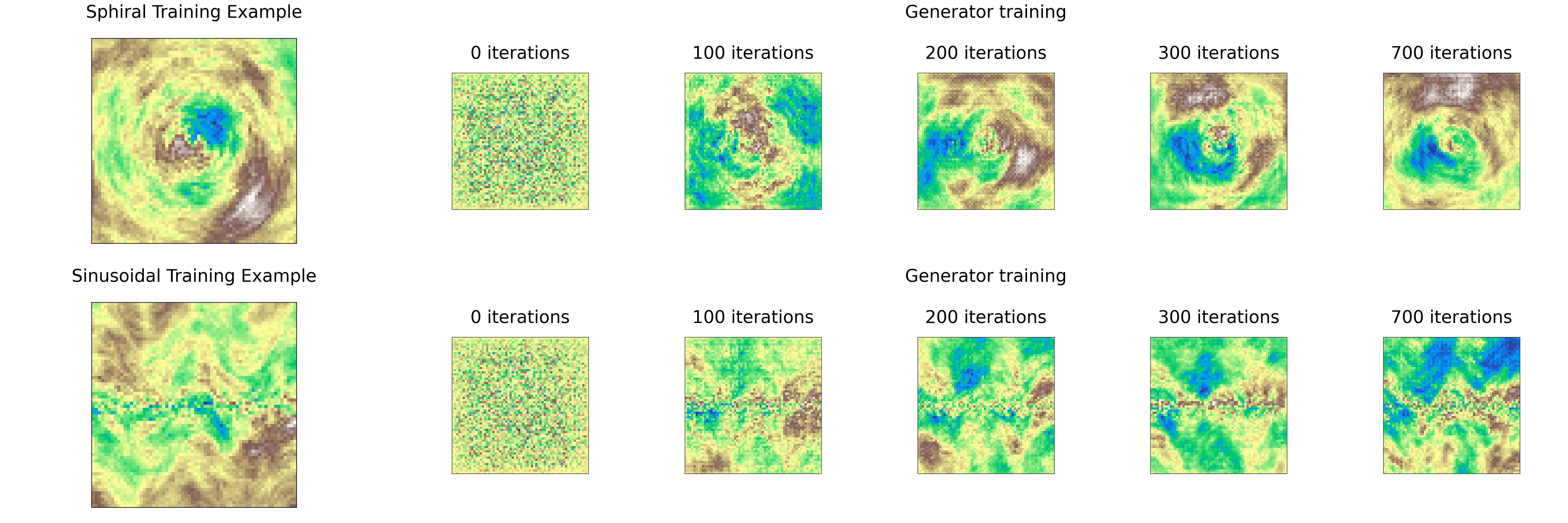}
    \caption{Comparison input synthetic data samples and generated samples from the GAN after some number of iterations for the spiral and sine wave structures.}
    \label{fig:2d_GAN_results}
\end{figure*}

Results comparing sample inputs to the GAN and the resulting generated unconditioned simulations can be observed in Figure \ref{fig:2d_GAN_results}. As shown, the initial outputs of the generator are pure noise. After around 100 iterations, the models begin to converge to generate samples comparable to their corresponding input images. Late iteration generated structures lie identical to the input training image structures, and the trained discriminator has trouble distinguishing between the two.

Using these trained networks, conditioned simulations for each of the models can be created. A set of known values at their respective coordinates are fed as supplementary information to the trained generator and discriminator models. The generator again tries to imitate realistic realizations, however now bound to the restriction of including the provided points. This allows us to generate plausible simulations of our known data that match the structure (orientation and ranges) of the input training data. The results of these conditioned 2D GAN simulations can be found in Figure \ref{fig:2d_gan_condition} below. Whereas these simulations can be swiftly generated, the conditioned non-stationary covariance simulations tend to yield better results than those generated by the GAN.

\begin{figure*}
    \centering
    \includegraphics[width=0.7\textwidth]{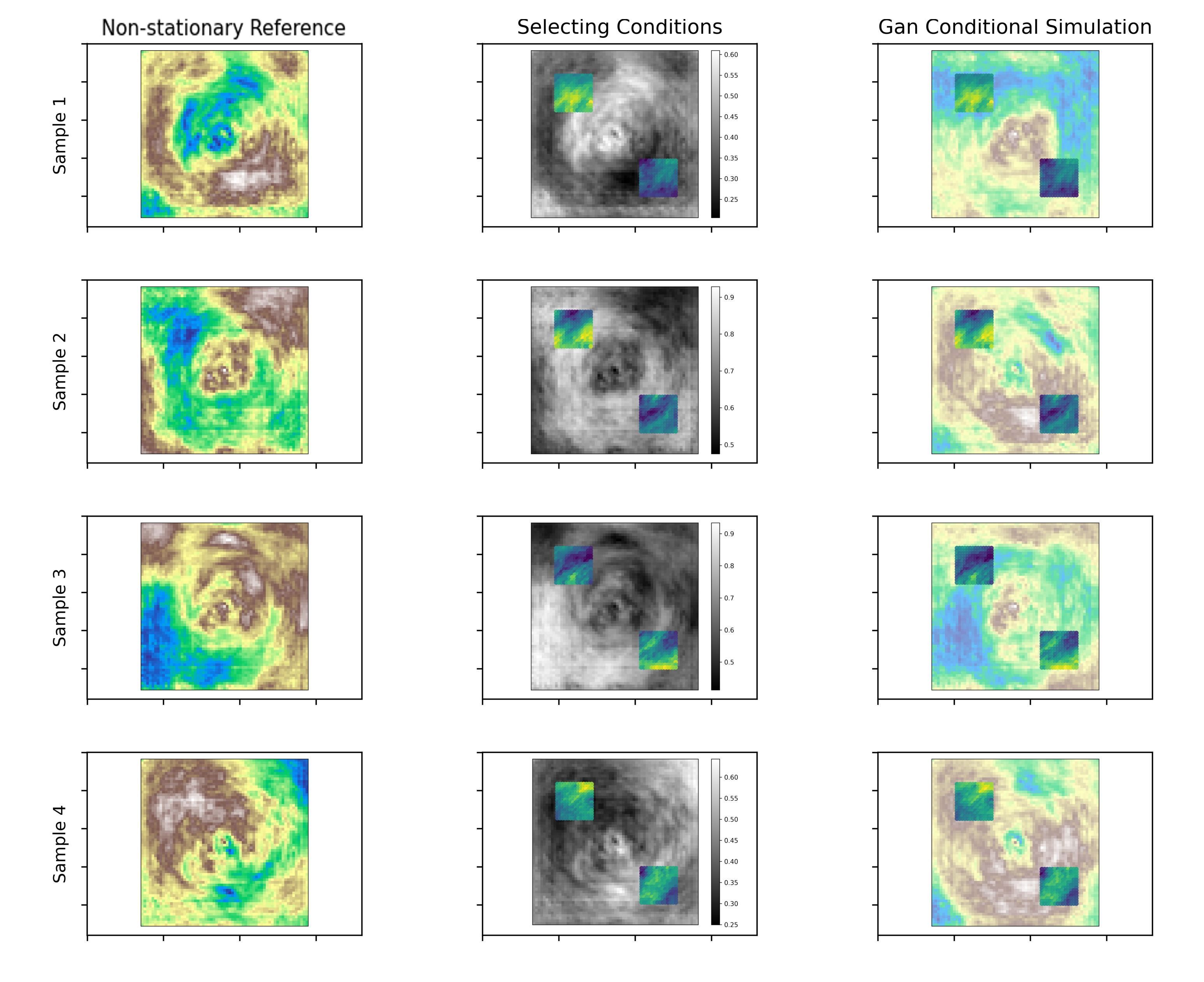}
    \caption{\textcolor{black}{First column: Sample unconditioned non-stationary simulations; Second Column: Selecting sub-regions of the reference simulations (greyscale) to be used as conditioning points for the 2D cGAN. The selected conditioning areas are visualized in color; Third Column: Output of the cGAN given the condition area. The output is visualized with a brighter color underneath the conditioned sub-regions.}}
    \label{fig:2d_gan_condition}
\end{figure*}

\begin{figure}
     \centering
     \begin{subfigure}[b]{0.5\textwidth}
        \centering
        \includegraphics[width=\textwidth]{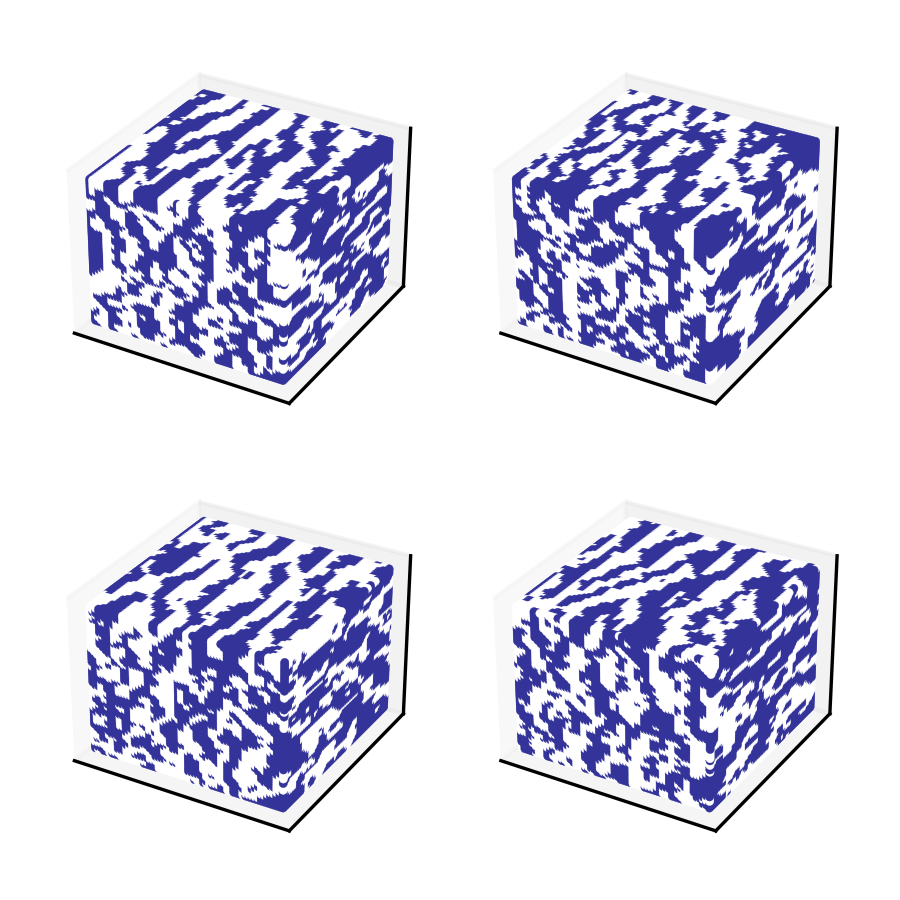}
        \caption{\textcolor{black}{Multipoint simulations, trained on the Maules Creek hydrofacies dataset, to be used as training samples for the 3D GAN.}}
        \label{fig:3d_GAN_input}
     \end{subfigure}
     \hfill
     \begin{subfigure}[b]{0.5\textwidth}
        \centering
        \includegraphics[width=\textwidth]{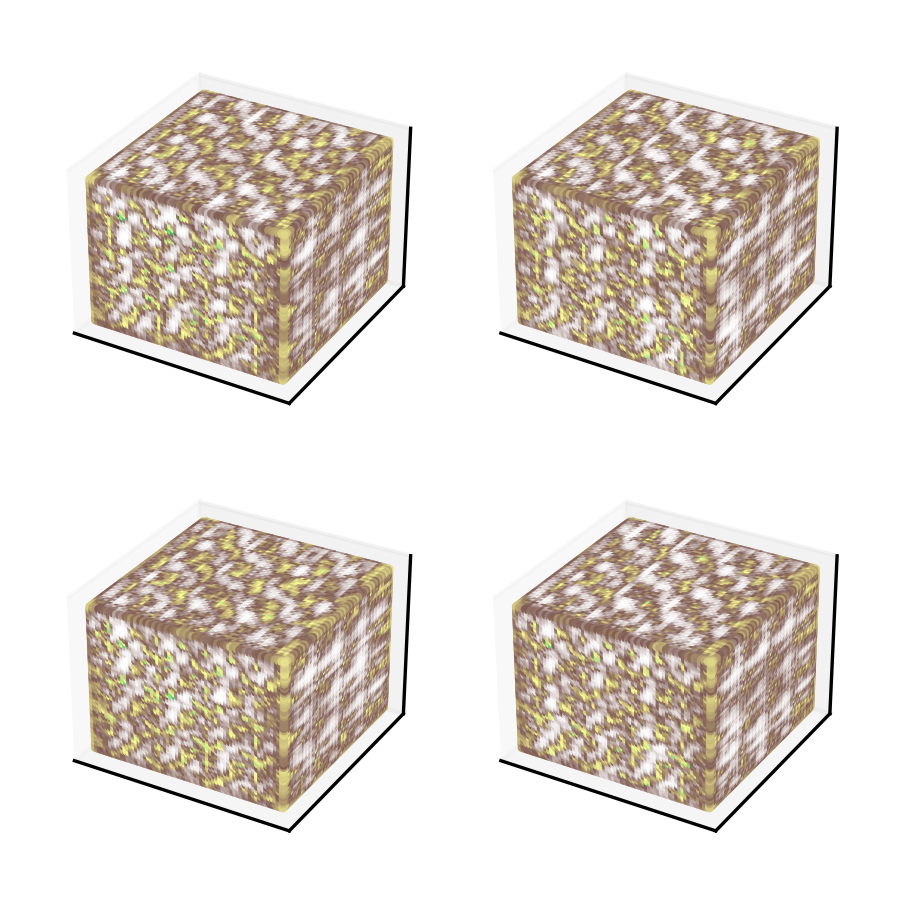}
        \caption{GAN results trained on the Maules Creek hydrofacies dataset after 60 iterations.}
        \label{fig:3d_GAN_results}
     \end{subfigure}
        \caption{Three dimensional GAN training results.}
        \label{fig:3d_GAN}
\end{figure}

To demonstrate an example for a 3D simulation, we utilise 1000 simulations generated from our multipoint method as training data for this model. Our observations for the two-dimensional GAN results remain consistent across the three dimensional samples as well, as shown in Figure \ref{fig:3d_GAN}. 

\section{Case Study: CVRHE}
\label{section:Case Study}

Building upon the previously described synthetic examples, estimation was conducted using real geochemical samples. This data comes from La Ceinture de roches vertes de la Haute-Eastmain (CRVHE), located in the Superior Province (approximately 320 km northeast of Chibougamau) and contains numerous gold deposits that were the subjects of intensive exploration work and brief gold production between 1994 and 1995 at the Eastmain mine. Several geological mapping and rock geochemistry sampling programs were undertaken since the late 1980's by the Ministère de l'Energie et des Ressources Naturelles (MERN) of Québec, Canada, resulting in the collection of more than 2,200 rock samples. These were mainly analyzed by a whole rock chemical method, which have allowed a precise petrological characterization of the rocks. The estimated variable for this case study is the normalized concentration of aluminum oxide ($Al_{2}O_{3}$). Figure \ref{fig:Ludo_data_input} below shows our input data for this dataset. 

In order to calculate non-stationary covariances on this data, a set of range and orientations for each point in the output grid must be provided. To select these values, magnetic response data from the same study area was analyzed, and any directional trend found was used in the corresponding location of the non-stationary input. Any areas where there is no clear trend in any direction was deemed isotropic (having equal range in all directions). This method gives us a reliable pillar on which to base our structure definitions as opposed to analyzing the input data manually. In the case of stationary covariance, we use the ranges 20km and 80km, with an angle of 45 degrees in the primary direction, as this structure is the most prominent through the study area. Our non-stationary target structure consists of two main areas with angles of 45 and 135 degrees, sharing ranges of 20km and 80km, with the majority of the area consisting of an isotropic structure with the range of 7km.

\begin{figure}
    \centering
    \includegraphics[width=.5\textwidth]{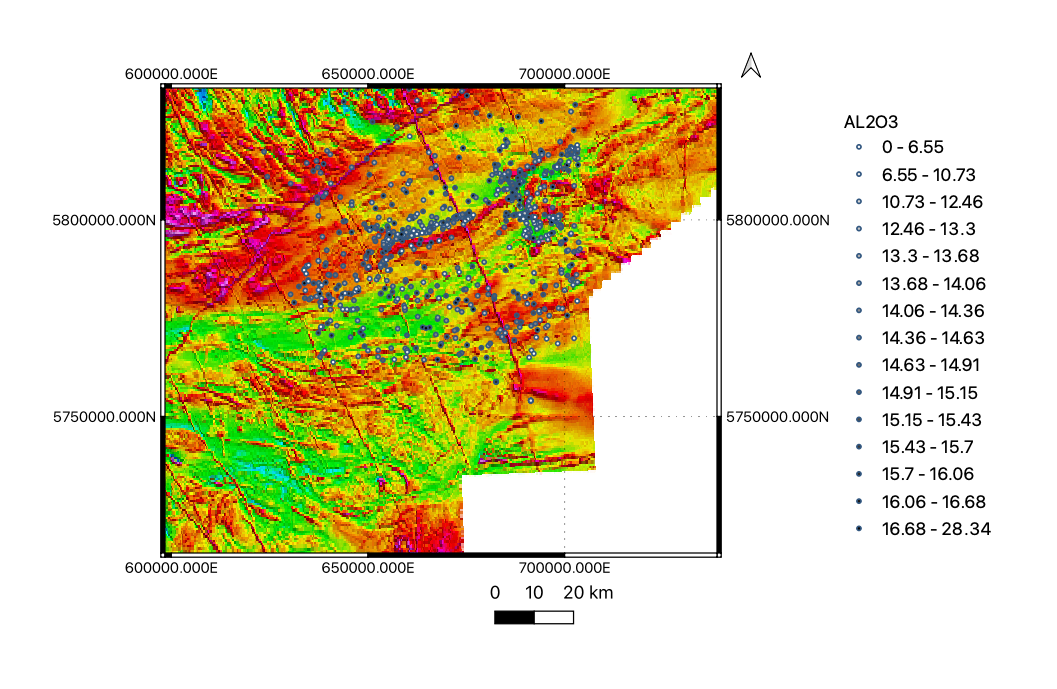}
    \caption{Aluminum oxide density data for non-stationary resource estimation and simulation. Points represent aluminum oxide density samples, and the colour map shows magnetic data.}
    \label{fig:Ludo_data_input}
\end{figure}

\begin{figure}
     \centering
     \begin{subfigure}[b]{0.4\textwidth}
         \centering
        \includegraphics[width=\textwidth]{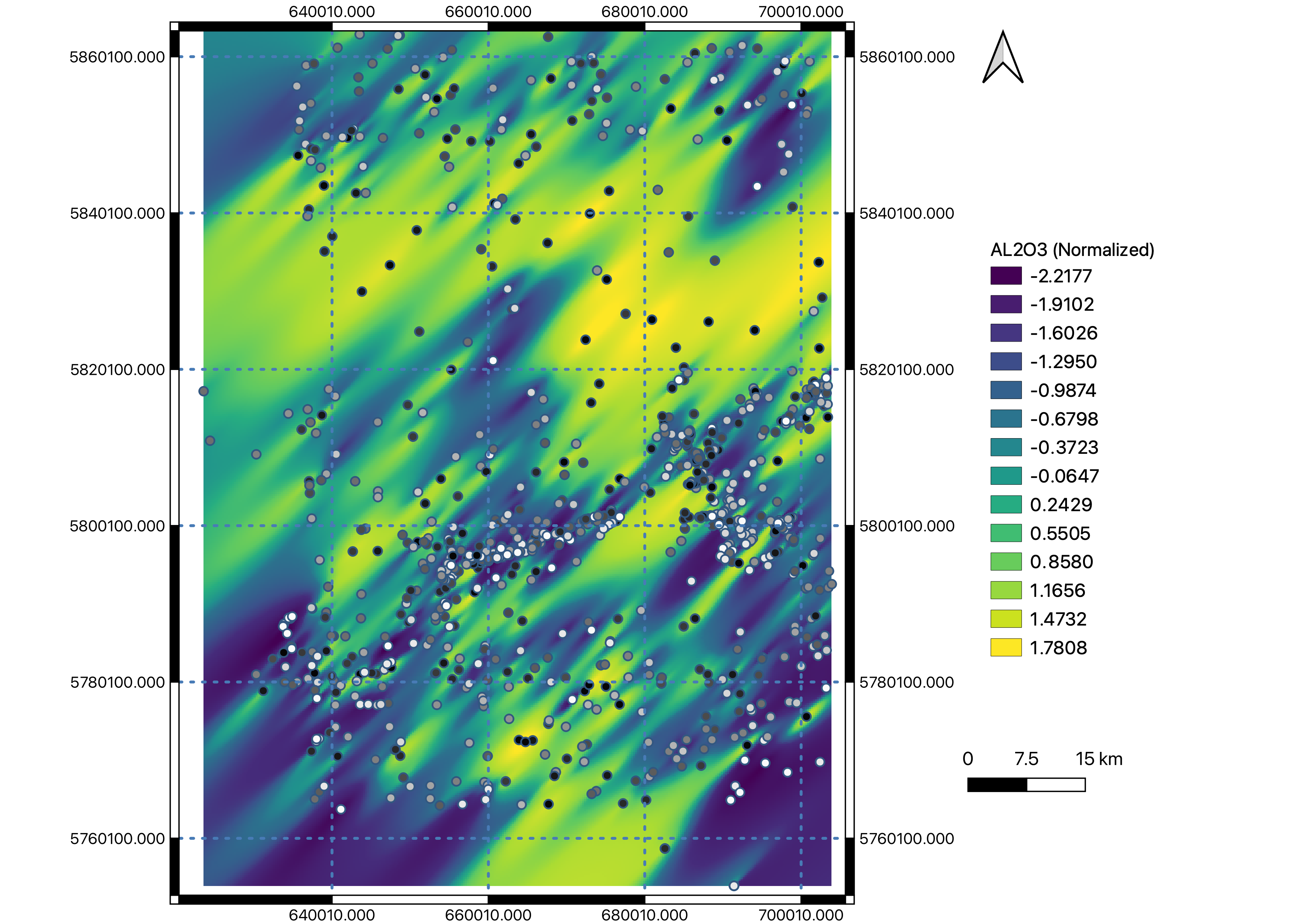}
        \caption{Stationary}
        \label{fig:Ludo_data_results}
     \end{subfigure}
     \hfill
     \begin{subfigure}[b]{0.4\textwidth}
         \centering
         \includegraphics[width=\textwidth]{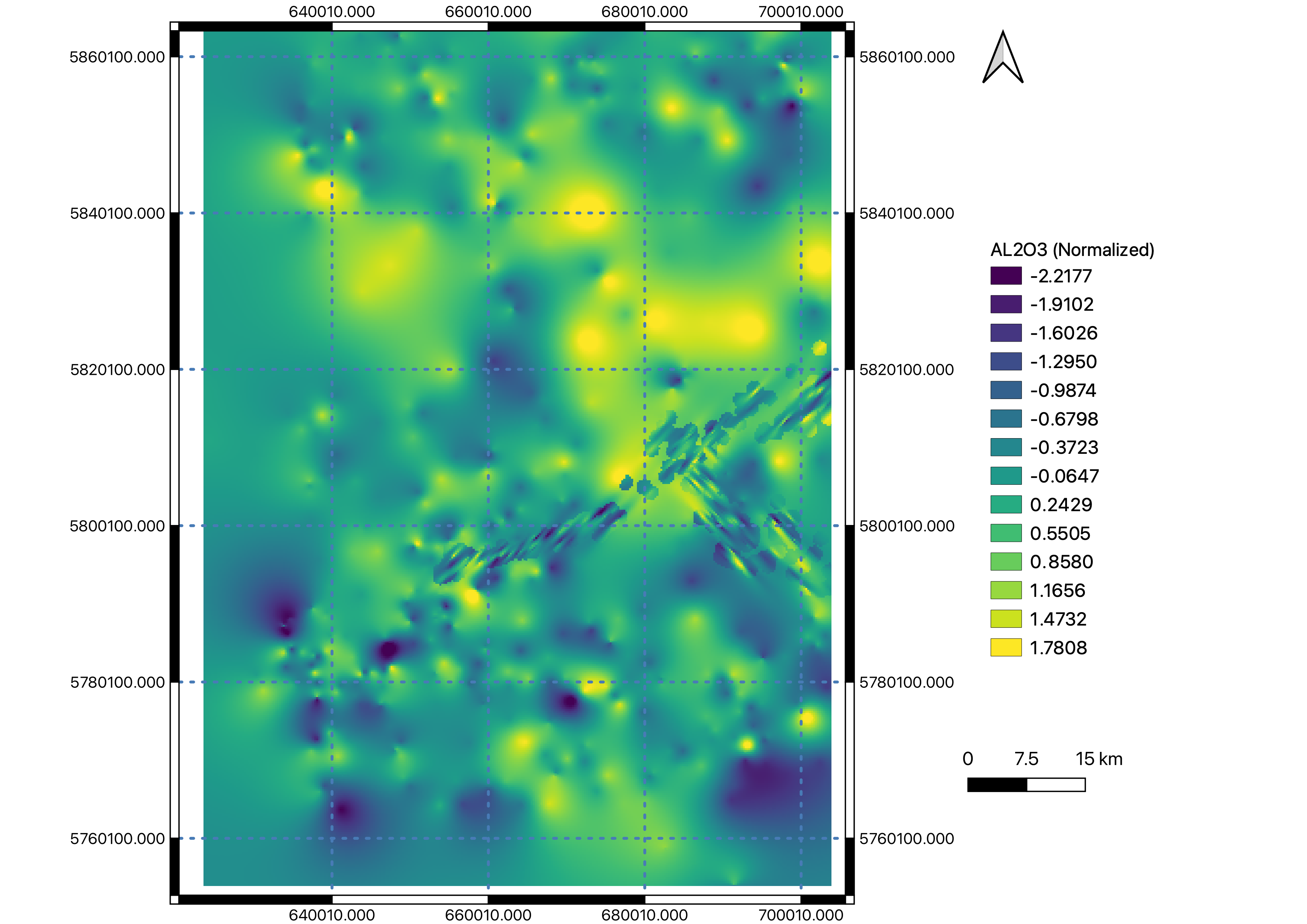}
         \caption{Non-Stationary}
         \label{fig:Ludo_data_results_nonstat}
     \end{subfigure}
        \caption{Result of aluminum oxide concentration simulation using stationary (a) and non-stationary (b) covariance calculations. }
        \label{fig:case_study_results}
\end{figure}

We calculate both stationary and non-stationary covariances in order to simulate and estimate surrounding aluminum oxide values, the results of which can be found in Figure \ref{fig:case_study_results}. For the stationary results, though areas that project the input structure show fair results, all other areas are greatly altered regardless of its similarity to the provided input. Our conditioned non-stationary results clearly highlight the two defined structures, while keeping the less dense portions of the study isotropic and independent of these two formations.

\section{Discussion and Conclusion}
\label{section:Discussion and Conclusion}
Admittedly, the quality of results obtained with the non-stationary covariance will largely rely on the realism of the initial geologic hypothesis underlying the parameterization of the non-stationary covariance. In the case of the spiral angle structure used in our non-stationary and GAN simulations, although this representation is not realistic in a geological setting, it is an effective stress test of the models. The results do, however, indicate a better performance of non-stationary over stationary covariance reflecting available geologic information. 

In a similar note, the input training images for our GAN simulations are in fact synthetically generated from our non-stationary and multipoint simulations. As such, it is plausible to presume that our results may change given real measurements as training data.

\textcolor{black}{The viability of variography, as well as other statistical moments, for SPDE and GAN has already been studied in previous works (\cite{LINDGREN2022100599}, \cite{chan2017parametrization}). Both simulation methods yield justifiable results when comparing the input samples and generated simulations. Rather than recovering exact variograms, inference of pattern statistics  and cost functions to measure similarity (either locally or globally) can also be applied \citep{coiffier20203d}.}



\begin{table*}[ht]
\centering
\resizebox{\textwidth}{!}{
\begin{tabular}{lrrrrrrrrrrrr}
\toprule
Frame size &    64  &    72  &    80  &     88  &     96  &     104 &     112 &     120 &     128 &     136 &      144 &      152 \\
\midrule
GAN-CPU preprocessing              &   0.13 &   0.17 &   0.19 &    0.25 &    0.28 &    0.31 &    0.35 &    0.42 &    0.45 &    0.52 &     0.57 &     0.64 \\
GAN-CPU training time (all epochs) &  71.80 &  48.67 &  65.82 &   65.65 &   92.52 &   88.58 &  116.50 &  114.38 &  177.90 &  150.47 &   192.64 &   185.86 \\
GAN-CPU prediction              &  12.22 &   4.31 &   7.67 &    6.13 &   11.97 &    8.49 &   14.17 &   10.87 &   24.65 &   13.57 &    21.97 &    16.51 \\
GAN-CPU total                   &  84.14 &  53.15 &  73.69 &   72.04 &  104.77 &   97.38 &  131.03 &  125.66 &  203.01 &  164.57 &   215.19 &   203.02 \\
GAN-GPU preprocessing              &   0.13 &   0.16 &   0.19 &    0.22 &    0.26 &    0.30 &    0.34 &    0.39 &    0.44 &    0.50 &     0.56 &     0.62 \\
GAN-GPU training time (all epochs) &   6.16 &   1.93 &   2.20 &    2.65 &    2.91 &    3.64 &    4.29 &    4.98 &    5.35 &    6.43 &     7.07 &     8.03 \\
GAN-GPU prediction              &   0.00 &   0.02 &   0.00 &    0.00 &    0.00 &    0.00 &    0.00 &    0.00 &    0.00 &    0.00 &     0.00 &     0.00 \\
GAN-GPU total                   &   6.30 &   2.11 &   2.40 &    2.87 &    3.18 &    3.94 &    4.64 &    5.37 &    5.79 &    6.92 &     7.63 &     8.65 \\
SPDE fitting                    &   0.01 &   0.01 &   0.01 &    0.01 &    0.01 &    0.02 &    0.03 &    0.03 &    0.04 &    0.04 &     0.05 &     0.05 \\
SPDE sampling                   &  18.68 &  25.60 &  32.97 &   41.47 &   51.12 &   94.62 &  113.63 &  134.52 &  156.93 &  180.22 &   204.34 &   237.64 \\
SPDE total                      &  18.69 &  25.61 &  32.98 &   41.48 &   51.13 &   94.63 &  113.66 &  134.55 &  156.97 &  180.26 &   204.38 &   237.69 \\
Non-stationary                   &  44.67 &  66.35 &  99.25 &  144.29 &  211.17 &  286.58 &  374.11 &  490.62 &  726.71 &  816.60 &  1020.14 &  1299.77 \\
\bottomrule
\toprule
Frame size &      160 &      168 &      176 &      184 &      192 &      200 &      208 &      216 &      224 &      232 &      240 &      248 \\
\midrule
GAN-CPU preprocessing              &     0.72 &     0.78 &     0.86 &     0.91 &     0.96 &     1.05 &     1.13 &     1.22 &     1.33 &     1.41 &     1.51 &     1.62 \\
GAN-CPU training time (all epochs) &   250.32 &   229.03 &   295.13 &   278.90 &   372.98 &   336.53 &   426.32 &   412.97 &   644.74 &   495.71 &   598.31 &   576.82 \\
GAN-CPU prediction              &    30.56 &    19.42 &    33.10 &    23.35 &    44.03 &    28.15 &    43.77 &    32.77 &    89.34 &    39.97 &    57.02 &    37.07 \\
GAN-CPU total                   &   281.60 &   249.23 &   329.09 &   303.16 &   417.97 &   365.74 &   471.22 &   446.96 &   735.42 &   537.09 &   656.85 &   615.51 \\
GAN-GPU preprocessing              &     0.68 &     0.75 &     0.81 &     0.89 &     0.98 &     1.05 &     1.14 &     1.23 &     1.32 &     1.41 &     1.51 &     1.61 \\
GAN-GPU training time (all epochs) &     9.85 &    10.91 &    12.00 &    13.57 &    14.83 &    17.79 &    19.31 &    21.29 &    23.07 &    26.17 &    28.21 &    33.90 \\
GAN-GPU prediction              &     0.00 &     0.00 &     0.00 &     0.00 &     0.00 &     0.00 &     0.00 &     0.00 &     0.00 &     0.00 &     0.00 &     0.00 \\
GAN-GPU total                   &    10.53 &    11.66 &    12.81 &    14.47 &    15.81 &    18.84 &    20.45 &    22.52 &    24.39 &    27.59 &    29.73 &    35.52 \\
SPDE fitting                    &     0.06 &     0.06 &     0.07 &     0.07 &     0.08 &     0.08 &     0.09 &     0.09 &     0.10 &     0.10 &     0.11 &     0.11 \\
SPDE sampling                   &   265.83 &   306.14 &   344.80 &   386.16 &   424.45 &   469.02 &   508.72 &   556.96 &   593.50 &   637.71 &   682.89 &   730.25 \\
SPDE total                      &   265.88 &   306.20 &   344.86 &   386.24 &   424.53 &   469.10 &   508.81 &   557.06 &   593.59 &   637.81 &   682.99 &   730.36 \\
Non-stationary                   &  1636.83 &  1907.63 &  2302.70 &  2786.91 &  3554.23 &  3816.60 &  4582.76 &  5421.46 &  6081.86 &  7200.04 &  8098.68 &  9470.60 \\
\bottomrule
\end{tabular}}
\caption{\textcolor{black}{Time measurements, in milliseconds, for the generation of one thousand 2D square simulations using GAN, SPDE and non-stationary methods.}} 
\label{table:time_table}
\end{table*}

In an experiment to compare the performances of different methods discussed in Section \ref{section:Methodology}, we tracked the time required to generate 1000 samples of spherical models for each of our methods as well as the benchmark SPDE method. This experiment was run using increasing frame sizes, between $64 \times 64$ to $248 \times 248$. A summary of the results of this experiment is provided in Table \ref{table:time_table}. 
\textcolor{black}{The sampling process for model has been broken down into its appropriate subprocesses to better convey the distribution of time throughout the simulation tasks. Our conditional GAN outperforms the non-stationary implementations, with total sampling time differences between the two methods growing exponentially as frame sizes increase.} 

\textcolor{black}{The results show that the non-stationary covariance simulations are the slowest to generate. Whereas GAN models can be pre-trained, non-stationary covariance calculations have no pre-processing that can be computed prior to generating samples. We also see that GAN simulations ran on the GPU are created faster than those using SPDE. GAN models ran on the CPU only exhibit shorter times at larger frame sizes. This difference in time exponentially grows with the size of the sample frames. Notably, the majority of the total time for GAN simulations come from the model training, with an insignificant amount of prediction time. Conversely, the majority of the time for the SPDE model comes from the sampling. This means using a pre-trained GAN, the time required to generate simulations would be a sliver of the total time shown.
Note that there are some caveats to this comparison. Namely, whereas the SPDE sampling time would diminish as the number of conditioning points increases. On the other hand, a GAN's prediction time will remain constant regardless of the number of condition samples. Also, every time the frame size or structure of the desired simulation were to change, the GAN would need to be retrained, significantly increasing the total time required as compared to only generating predictions.}

\textcolor{black}{Besides the improved simulation time for a GAN, it provides more flexibility in parameterization. Alongside direct samples or conditioning points, it is possible to setup a GAN such that it may accept a wide variety of engineered features to help define the output structure. Vectorized representations for each sample can be generated in the latent space, which can in turn be used alter the output simulation.}

Stochastic simulation methods have historically focused on stationary, or single-structured simulation methods, though due to advances in computing hardware, is it currently viable to apply non-stationary covariance, greatly improving simulation quality. The use of machine learning models can also be used to effectively simulate non-stationary structures. To this avail, this work implements non-stationary covariance, multipoint and GAN simulations using synthetic input and training images, and compares the performance of these methods to the benchmark SPDE technique. A case study was conducted by applying non-station covariance methods to real geochemical data, generating realistic simulated data following the provided structures. 

In the near future, we expect to see more developments using deep learning techniques such as GAN-inspired methods for geostatistical simulation and inversion of geophysical data \citep{laloy2018training}. 

\section{Acknowledgements}
The authors would like to thank Ludovic Bigot for providing the geochemical data which is used for our case study. We would also like to thank Grace Dupuis for her contribution to this work.

\clearpage
\bibliographystyle{cas-model2-names.bst}
\bibliography{nonstat.bib}

\end{document}